\documentclass[aps,prl,twocolumn,floatfix,superscriptaddress]{revtex4}

\usepackage{amssymb,amsmath,amstext}                
\usepackage{graphicx}   
\usepackage{epstopdf}                                               
\usepackage{color}                                                     
\usepackage{bm}                                                        
\usepackage{appendix}          

\usepackage[utf8]{inputenc}
\usepackage{lipsum}

\usepackage{ulem}
\usepackage{latexsym}
\usepackage[colorlinks=true,citecolor=blue,linkcolor=magenta]{hyperref}
\usepackage{soul}
\usepackage{amsmath}
\usepackage{float}
\DeclareMathOperator{\sign}{sign}

\def\be{\begin{equation}}
\def\ee{\end{equation}}
\def\bea{\begin{eqnarray}}
\def\eea{\end{eqnarray}}
\def\ra{\rangle}
\def\la{\langle}
\def\bi{\begin{itemize}}
\def\ei{\end{itemize}}

\definecolor{dgreen} {RGB}{78,138,21}

\definecolor{giergiel} {RGB}{0,168,128}

\definecolor{purple} {RGB}{128,0,160}

\begin{document} 
\title{Inseparable time-crystal geometries on the M\"obius strip}
\author{Krzysztof Giergiel} 
\affiliation{
Instytut Fizyki Teoretycznej,
Uniwersytet Jagiello\'nski, ulica Profesora Stanis\l{}awa \L{}ojasiewicza 11, PL-30-348 Krak\'ow, Poland}
\author{Arkadiusz Kuro\'s} 
\affiliation{
Instytut Fizyki Teoretycznej,
Uniwersytet Jagiello\'nski, ulica Profesora Stanis\l{}awa \L{}ojasiewicza 11, PL-30-348 Krak\'ow, Poland}
\author{Arkadiusz Kosior}
\affiliation{
Instytut Fizyki Teoretycznej,
Uniwersytet Jagiello\'nski, ulica Profesora Stanis\l{}awa \L{}ojasiewicza 11, PL-30-348 Krak\'ow, Poland}
\affiliation{Max-Planck-Institut f\"ur Physik Komplexer Systeme,
N\"othnitzer Strasse 38, D-01187, Dresden, Germany}
\author{Krzysztof Sacha} 
\affiliation{
Instytut Fizyki Teoretycznej,
Uniwersytet Jagiello\'nski, ulica Profesora Stanis\l{}awa \L{}ojasiewicza 11, PL-30-348 Krak\'ow, Poland}

\begin{abstract}
Description of periodically and resonantly driven quantum systems can lead to solid state models where condensed matter phenomena can be investigated in time lattices formed by periodically evolving Wannier-like states. Here, we show that inseparable two-dimensional time lattices with the M\"obius strip geometry can be realized for ultra-cold atoms bouncing between two periodically oscillating mirrors. Effective interactions between atoms loaded to a lattice can be long-ranged and can be controlled experimentally. As a specific example we show how to realize a Lieb lattice model with a flat band and how to control long-range hopping of pairs of atoms in the model.
\end{abstract}
\date{\today}

\maketitle
\textbf{Introduction.}
In the last few decades engineering of elaborate optical potentials has been a prominent subject of both theoretical and experimental research in ultra-cold atoms \cite{lewenstein2017ultracold,Eckardt2017}. Recent experimental techniques enable not only creation of periodic optical potentials of various geometries \cite{Tarruell2012,windpassinger2013engineering}, but also manipulation of parameters of the effective models and introduction of artificial gauge fields \cite{Goldman2014}. The later allows to realize topologically non-trivial energy bands which is the cornerstone of topological insulators and quantum Hall systems \cite{Hasan2010}. The real space topology proves to be equally important, for example, it has been shown that global properties of spinless particles on the M\"obius ladder can be locally described by a non-Abelian gauge potential \cite{Guo2009,Zhao2009}
or that the quantum Hall effect is forbidden on non-orientable surfaces \cite{Beugeling2014}. 
Unfortunately, realization of non-trivial real space topologies can be challenging. Although it has been shown that topologically non-trivial one dimensional ladder geometries can be implemented by using a synthetic dimension \cite{Boada2012,boada2015quantum}, higher dimensional systems have remained elusive so far. 

On the other hand, recently there has been an increasing number of theoretical works on time crystals \cite{Wilczek2012,Sacha2015,Khemani16,ElseFTC,
Yao2017,Lazarides2017,
Russomanno2017,Ho2017,
Huang2017,Iemini2017,Wang2017,
Zeng2017,Surace2018,
Mizuta2018,Giergiel2018a,
Kosior2018,Kosior2018a,
Pizzi2019a,liang2018floquet,
Bomantara2018,Fan2019,Kozin2019,
Matus2019,Pizzi2021,
SyrwidKosiorSacha2020,Syrwid2020,
Russomanno2020,giergiel2020creating,Wang2020,kuros2020}, followed by experimental demonstrations \cite{Zhang2017,Choi2017,Pal2018,Rovny2018,Autti2018,
Kreil2018,Rovny2018a,
Smits2018,Liao2018,Autti2021}, and modelling of crystalline structures in periodically driven systems \cite{Guo2013,sacha16,Mierzejewski2017,Lustig2018,Giergiel2018,Giergiel2018c} (for reviews see  \cite{Sacha2017rev,khemani2019brief,guo2020,SachaTC2020}). The later opens a path to realisation of temporal analogs of condensed matter physics and exploration of novel phenomena present exclusively in the time dimension. 
In particular, in this Letter we show the construction of two dimensional insepearable time lattices  which naturally entails the M\"obius strip geometry.  Specifically, we identify reduction of the description of atoms resonantly bouncing between two periodically oscillating mirrors to the tight-binding Hamiltonian, where particles can tunnel between localized Wannier-like wave-packets which evolve periodically along classical resonant trajectories. The crystalline structure corresponding to the tight-binding Hamiltonian can be observed not in space but in the time domain. That is, if we locate a particle detector close to a resonant trajectory, the dependence of the probability of clicking of the detector as a function of time reproduces a cut of the crystalline structure described by the model \cite{SachaTC2020}. This reflects the fact that in time crystals the roles of time and space are interchanged. 

In the following we show how to realize tight binding models on a two dimensional (2D) crystalline structure on the M\"obius strip in the time domain. We propose a universal setup where the emergent lattice geometry can be shaped almost at will depending on the driving protocol of the mirrors. As a particular example we choose the Lieb lattice with a flat band \cite{Taie2015,Dauphin2016,Leykam2018,Tylutki2018,Taie2020} where dynamics of atoms is governed solely by interactions. We stress that the effective interactions of model are long-ranged and can be experimentally controlled. This creates a unique platform to study exotic flat band many-body physics. In the next sections we describe the main elements of the theoretical approach, leaving the details in \cite{Sup1}.

\begin{figure}[h!] 	            
\includegraphics[width=.8\columnwidth]{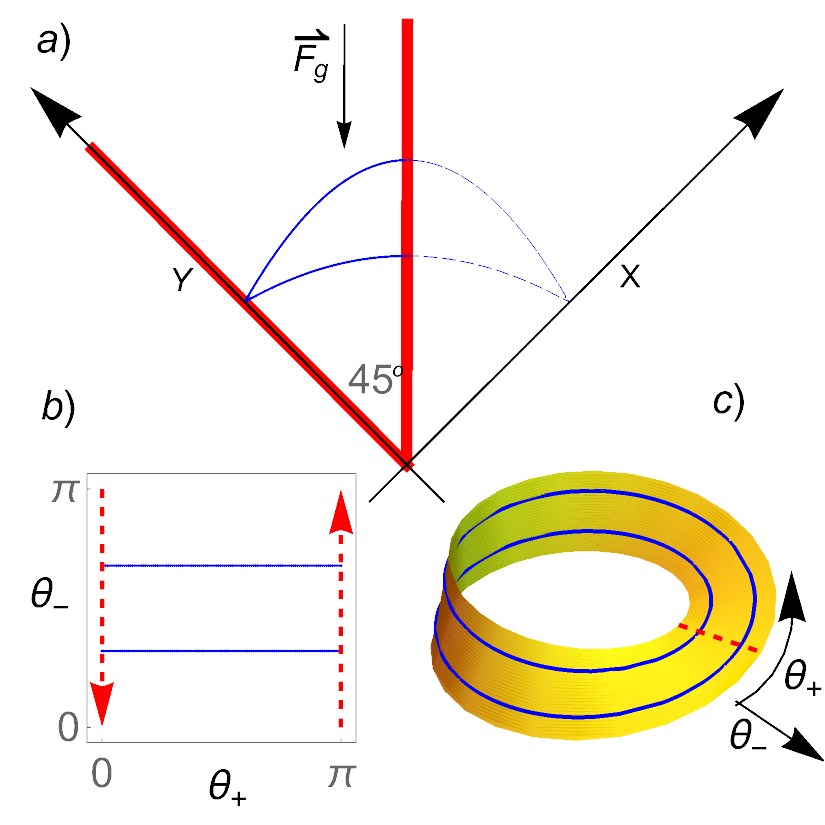} 
\caption{(a) A geometry of the system where a particle in the presence of the gravitational force $\vec F_g$ is bouncing between two mirrors (thick red lines) forming a $45^\circ$ wedge.
 (b) If the mirrors do not oscillate, a set of trajectories (a sample trajectory shown in blue) corresponding to equal energies $E_x=E_y$ cover a region with  $ \theta_{\pm}\in [0,\pi)$. In a collision with the vertical mirror, i.e. at $\theta_{+}=\pi$, the momenta components of a particle are exchanged what actually reverses the direction of the momentum vector $p_{x,y}\rightarrow -p_{x,y}$ because for $E_x=E_y$ we have $p_x=-p_y$. This results in $\theta_{\pm}\rightarrow \pi-\theta_{\pm}$. These conditions identify points $\{\theta_{+}=\pi,\theta_{-}\}=\{\theta_{+}=0,\pi-\theta_{-}\}$ and actually define the M\"obius strip geometry (c).}
\label{mirrors}   
\end{figure}

{\bf  M\"obius strip geometry.}
Let us start with a classical particle bouncing between two static mirrors located at $x=0$ and $x-y=0$, which form a wedge with the angle $45^\circ$ (Fig.~\ref{mirrors}). In the gravitational units \cite{Buchleitner2002,footnote1}, the Hamiltonian reads
$H_0=(p_x^2+p_y^2)/2+x+y$ with the constraint $y\ge x\ge 0$
coming from the hard wall potential of the mirrors (for a Gaussian shaped  mirror potential see \cite{giergiel2020creating}).
When a particle collides with the vertical mirror, its  momenta are exchanged $p_x \leftrightarrows p_y$, whereas when a particle hits the other mirror, $p_y$ remains the same but $p_x\rightarrow-p_x$, see Fig.~\ref{mirrors}.

To find how to describe a particle confined in the wedge with the angle $45^\circ$ one can start with the problem of two perpendicular mirrors. When the angle between two mirrors is $90^\circ$, the system is separable in the Cartesian coordinate frame \cite{Richter1990,wojtkowski1990system} and it is convenient to switch to the action-angle variables $I_{\alpha}$ and $\theta_{\alpha}$ with $\alpha=x,y$.  Then, the Hamiltonian $H_0$ depends on the actions $I_{\alpha}$ only \cite{landau1982mechanics,Lichtenberg1992}. The dynamics of the angles is given by Hamilton's equations $\dot \theta_{\alpha} = \partial H_0/\partial I_{\alpha} \equiv \Omega_{\alpha} (I_\alpha)$, where $\Omega_{\alpha}(I_\alpha)$ are frequencies of motion along the $x$ and $y$ directions. Since the actions $I_{\alpha}$ are constants of motion, the solution for the angles are trivial, $\theta_{\alpha}(t)=  \Omega_{\alpha}(I_\alpha) t + \theta_{\alpha}(0) \, (\mbox{mod}\, 2\pi)$. Motion of a particle is confined on a surface of a two-dimensional torus. 
In this Letter we consider periodic trajectories of a particle which are symmetric with respect to the vertical mirror. It implies that the initial conditions correspond to equal energies of the $x$ and $y$ degrees of freedom, i.e. $E_x=E_y$ (or $I_x=I_y$) and thus $\Omega_x(I_x)=\Omega_y(I_y)$. To reduce the number of frequencies we perform a canonical transformation from $(I_{\alpha},\theta_{\alpha})$ to new variables $I_{\pm}=I_y\pm I_x$ and $\theta_{\pm}=(\theta_y\pm \theta_x)/2$ \cite{Sup1}. The equations of motion in such variables have the form $\dot{I}_{\pm}=0$, $\dot{\theta}_{-}=0$ and $\dot{\theta}_{+}=\Omega_{+}(I_{+})$ where $I_{-}=0$ and the value of the action $I_{+}$ determines the frequency of a periodic orbit \cite{antonowicz1981freedom}. Thus, $\theta_{+}(t)=\Omega_+(I_+)t+\theta_+(0)$ describes motion along a periodic orbit while $\theta_{-}$ is a constant. 

Let us come back to the wedge with the angle $45^\circ$, where the motion is restrained to $y\ge x$ (or equivalently $0<\theta_{+}\leq \pi $). When a particle bounces off a vertical mirror,  the  momenta are exchanged $p_x \leftrightarrows p_y$. For  $E_x=E_y$, we have $p_x=-p_y$ and therefore  $p_{\alpha} \rightarrow -p_{\alpha}$ at $y=x$, or in other words  $\theta_{\pm} \rightarrow \pi - \theta_{\pm}$ at $\theta_{+}=\pi$. The latter identifies points $\{\theta_{+}=\pi,\theta_{-}\}=\{\theta_{+}=0,\pi-\theta_{-}\}$ and actually defines the M\"obius strip geometry (see Fig.~\ref{mirrors}). In order to realize condensed matter physics on the M\"obius strip, oscillations of the mirrors will be turned on. We will see that resonant bouncing of a single atom or a cloud of atoms between the oscillating mirrors can be described by solid state models. The emerging crystalline structures will be observed not in space but in the time domain.

{\bf Oscillating mirrors.}
Let us assume that the mirror located around $x=0$ oscillates in time like $f_x(t)=-(\lambda_1/\omega^2)\cos(\omega t)-(\lambda_2/4\omega^2)\cos(2\omega t)$ while the vertical one like $f_{y-x}(t)=(\lambda_3/4\omega^2)\cos(2\omega t+\phi)$ where $\lambda_{1,2,3}$ are amplitudes and $\phi$ is a constant phase. It is convenient to switch to the frame oscillating with the mirrors. Then, the mirrors are static and the Hamiltonian of an atom reads $H=H_0+(x+y)f_x''(t)+yf_{y-x}''(t)$, see  \cite{Sup1}. 
We focus on the resonant driving of an atom where the frequency $\omega$ of the oscillations of the mirrors fulfills the $s:1$ resonant condition, i.e. $\omega=s\Omega_+(I_{+}^{0})$ where $s$ is an integer number, $I_{+}^{0}$ is the resonant value of the action $I_+$ and $I_-=I_{-}^{0}=0$. 

\begin{figure}[t!]                 
\includegraphics[width=.95\columnwidth]{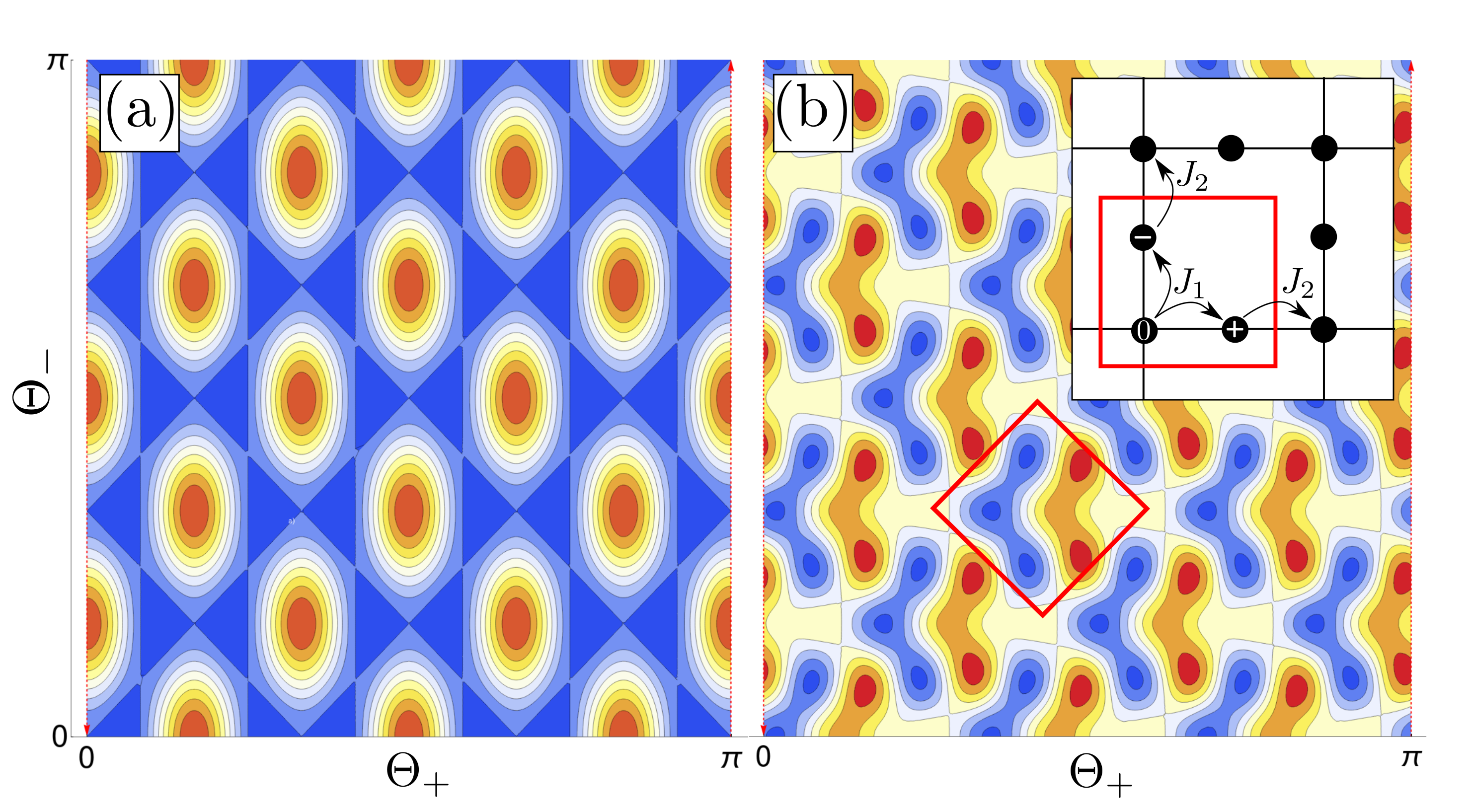}
\caption{Examples of the effective potential in Eq.~(\ref{heff}).
 Dark blue color represents areas around maxima of the effective potential which correspond to the lowest energies of a particle with a negative effective mass. The geometry of the $\{\Theta_+,\Theta_-\}$ space is the M\"obius strip geometry as in Fig.~\ref{mirrors}. (a): The effective potential for $\lambda_3/\lambda_1=4$, $\lambda_2=0$ and $\phi=0$ creates a honeycomb lattice structure. (b): Maxima of the effective potential for $\lambda_2/\lambda_1=4$, $\lambda_3/\lambda_2=1.62$ 
 and $\phi=\pi/4$ correspond to the the Lieb lattice with a well separated central flat band. A unit cell (red square) of the Lieb lattice is composed of three sites. Inset: A tunneling structure in the Lieb lattice. }
\label{eff_pot}
\end{figure}

In order to describe classical motion of an atom close to resonant trajectories, one may apply the secular approximation approach which in the action-angle variables and in the moving frame, $\Theta_{+}=\theta_{+}-\omega t/s$ and $\Theta_{-}=\theta_{-}$, leads to the following effective Hamiltonian \cite{Sup1} 
\be
\begin{split}
H_{\rm eff}=-\frac{P_{-}^2+P_{+}^2}{2|m_{\rm eff}|}- \frac{\lambda_2}{2\omega^2} \cos \left(2s\Theta_{+} \right) \cos \left(2s\Theta_{-}\right) \\
- \frac{2\lambda_1}{\omega^2}\cos(s\Theta_{+})\cos(s\Theta_{-}) +\frac{\lambda_3}{4 \omega^2}\cos \left(2s\Theta_{+}+\phi\right), 
\label{heff}
\end{split}
\ee 
where $P_{\pm}=I_{\pm}-I_{\pm}^{0}$ and $|m_{\rm eff}|=(3I_{0+})^{4/3}/(2\pi^2)^{1/3}$. The  Hamiltonian (\ref{heff}) describes a particle with the negative effective mass $-|m_{\rm eff}|$ in the presence of an inseparable lattice potential which is moving on the M\"obius strip because at $\Theta_{+}=\pi$ there are the flips ${\Theta_{\pm}} \rightarrow \pi- \Theta_{\pm}$.
Different parameters of the mirrors' oscillations allow one to realize different crystalline structures of the effective potential in (\ref{heff}). For example for $\lambda_3/\lambda_1=4$, $\lambda_2=0$ and $\phi=0$, a honeycomb lattice \cite{windpassinger2013engineering,Tarruell2012} can be realized [Fig.~\ref{eff_pot}(a)] while for $\lambda_2/\lambda_1=4$, $\lambda_3/\lambda_2=1.62$ and $\phi=\pi/4$, the Lieb lattice with a flat band emerges [Fig.~\ref{eff_pot}(b)]. In the following we focus on the Lieb lattice case as a concrete example.

To obtain a quantum description of a particle resonantly bouncing between the mirrors one can either quantize the classical Hamiltonian (\ref{heff}), i.e. replace $P_{\pm}\rightarrow-i\partial/\partial\Theta_{\pm}$, or apply the fully quantum secular approximation method for the Floquet Hamiltonian $H_F=H-i\partial_t$ (see \cite{Sup1}). The former is very useful to understand what kind of the effective behavior we can expect. The latter is a more systematic quantum description which allows one to easily incorporate the boundary conditions on the mirrors and particle interactions and we use it to obtain all quantum results of the Letter. These two quantum approaches agree very well with each other if $I_{+}^{0}\gg 1$.

We concentrate on an example where the effective potential in the Hamiltonian (\ref{heff}) correspond to the Lieb lattice [Fig.~\ref{eff_pot}(b)]. The Lieb lattice is a Bravais lattice with a three point basis, and therefore the lattice sites can be labeled by a unit cell index $j$ and an intra cell index $\beta=0,\pm$, see Fig.~\ref{eff_pot}(b). Description of the lowest energy manifold of the effective Hamiltonian can be reduced to the tight-binding model 
 \be\label{H_lieb}
 H_{F} \approx -J_1 \sum_{i,\beta=\pm}\hat a_{i,0}^\dagger \hat a_{i,\beta}-J_2 \sum_{\la ij\ra,\beta=\pm}\hat a_{i,0}^\dagger \hat a_{j,\beta} +\mbox{H.c.}
 \ee 
where $\hat a_{i,\beta}/\hat a_{i,\beta}^\dagger$ are bosonic operators that annihilate/create a particle in the Wannier states  $W_{i,\beta}(\Theta_+,\Theta_-)$. $J_{1}$ and $J_2$ are intra- and intercell tunneling amplitudes respectively, cf. Fig.~\ref{eff_pot}(b). 
As long as $J_1\ne J_2$, eigenvalues of Eq.~\eqref{H_lieb} form  three separated bands, where the central one is flat \cite{Taie2020,Sup1}. In the flat band the group velocity is zero and consequently the transport in the flat band is totally ceased unless we deal with a many-body system with interactions.

The Hamiltonian (\ref{heff}) indicates that in the moving frame we deal with a crystalline structure in the $\{\Theta_+,\Theta_-\}$ space. In the tight-binding approximation (\ref{H_lieb}) eigenstates of an atom are superposition of the Wannier states, $\psi(\Theta_+,\Theta_-)=\sum_{i,\beta}c_{i,\beta}W_{i,\beta}(\Theta_+,\Theta_-)$. When we return to the laboratory frame, no crystalline structure is observed in the Cartesian coordinates $x$ and $y$. However, if a detector is located close to a resonant trajectory (i.e. we fix $\theta_+$ and $\theta_-$ and $I_\pm\approx I_{\pm}^0$), then the dependence of the probability of clicking of the detector as a function of time reproduces a cut of the probability density in the $\{\Theta_+,\Theta_-\}$ space, i.e. $|\psi(\Theta_+,\Theta_-)|^2=|\psi(\theta_+-\omega t/s,\theta_-)|^2$. Different locations of the detector (different $\theta_\pm$) correspond to different cuts of the crystalline structure in the $\{\Theta_+,\Theta_-\}$ space. Note, that such a crystalline structure in time is not a result of spontaneous breaking of time translation symmetry. It is a time lattice which emerges in the dynamics of the system due to the external driving similarly like in the case of photonic crystals which do not form spontaneously because periodic modulation of the refractive index in space has to be imposed externally. 

\begin{figure*}
\includegraphics[width=1\textwidth]{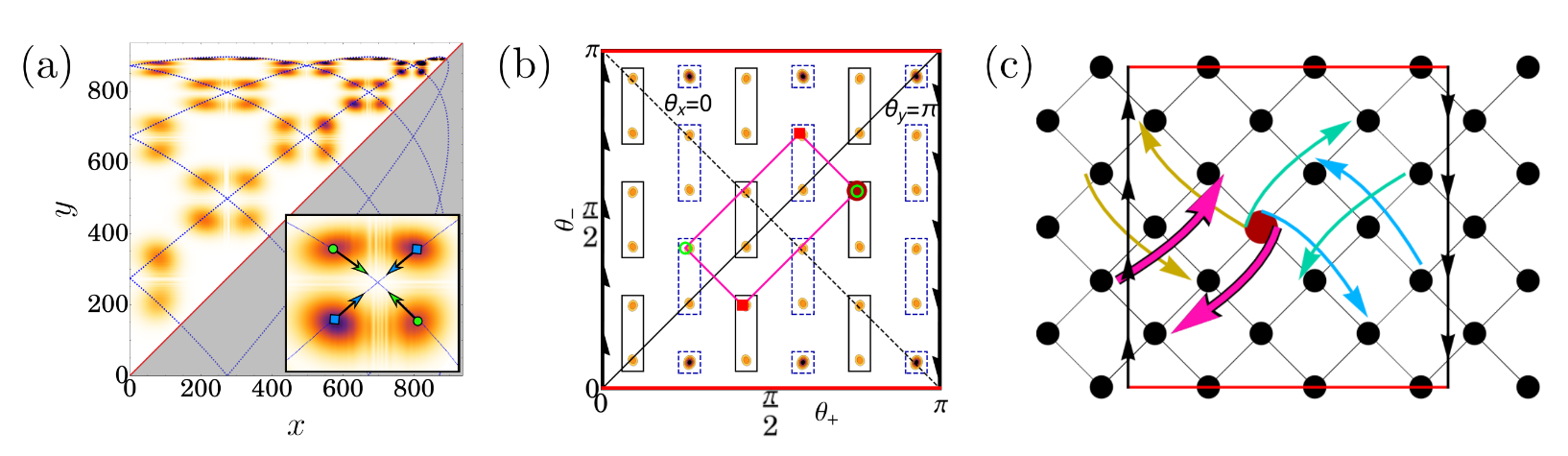} 
\caption{(a): Probability density (in the lab frame and in the Cartesian coordinates at $t=\omega\pi/5$) of the  Wannier states $w_i$ belonging to the flat band of the effective Lieb lattice potential, cf. Fig.~\ref{eff_pot}(b). Inset: A zoom on four encountering localized wavepackets belonging to four different Wannier states $w_i$, $w_j$, $w_k$, $w_l$. 
(b): Same as in (a) but in the   $\{\theta_+,\theta_-\}$ space. The Wannier states, enclosed by rectangles, are either superpositions of two localized wavepackets or just a single one at the edge of the M\"obius strip. In the course of time evolution the entire structure is moving uniformly along the $\theta_+$ axis and fulfills the M\"obius strip boundary conditions. 
 (c): The hopping structure of the effective lattice of the flat band, where black dots correspond to the Wannier states $w_i$, and arrows of the same color indicate hoppings of atomic pairs. The horizontal direction of the lattice is related to the direction along the M\"obius strip, cf. Fig.~\ref{mirrors}(c). Note that for illustrative purposes we have only shown the hoppings along the smallest symmetric retangles [cf. panel (b)] that invole anihilation of one atom in a central (brown) site. Panels correspond to $s=6$, $\omega=0.315$, $\lambda_1=2.48\cdot 10^{-4}$, $\lambda_2=9.9\cdot 10^{-4}$, $\lambda_3=1.61\cdot 10^{-3}$ and $\phi=\pi/4$ in Eq.~\eqref{heff}.  
}
\label{regale}
\end{figure*}


{\bf Quantum many-body physics in the flat band.} 
In the previous paragraphs we have shown how to realize an effective potential in the $\{\Theta_+,\Theta_-\}$ space, where a localized particle tunnels between the Wannier states $W_{j,\beta}(\Theta_+,\Theta_-)$ centered at the sites of the Lieb lattice [Eq.~\eqref{H_lieb}]. The eigenstates of the flat band can be chosen as the maximally localized Wannier states $w_j$. For $J_1/J_2\gg 1$, the Wannier states $w_{j}$ spanning the flat band can be approximated by superpositions of two localized wave-packets  $w_j \approx (W_{j,+}-W_{j,-})/\sqrt{2}$ for the bulk states or $w_j\approx W_{j,\pm}$ for the states close to the edge of the M\"obius strip \cite{Sup4}, see Fig.~\ref{regale}. 

Hopping of bosons in the flat band can only happen if there are interactions between them. In ultra-cold atoms, the interactions are zero-range and we assume that interaction energy per particle is much smaller than the energy gaps between the flat and adjacent bands. Then, we may still restrict to the flat band only and the effective many-body Floquet Hamiltonain reads \cite{SachaTC2020} 

\bea\label{H_fband}
 H_F&=&\frac{1}{sT}\int_0^{sT}dt\int dxdy\;\hat\psi^\dagger\left(H-i\partial_t+\frac{g_0}{2}\hat\psi^\dagger\hat\psi\right)\hat\psi
\cr &\approx&\sum_{ijkl}U_{ijkl}\hat b^\dagger_i\hat b_j^\dagger\hat b_k\hat b_l+{\rm const},
\eea
where $H$ is the single particle Hamiltonian, $\hat\psi\approx\sum_{i=1}^{s(s+1)/2}w_{i}\hat b_i$ with the bosonic operators $[\hat b_i,\hat b_j^\dagger]=\delta_{ij}$, and $U_{ijkl}=(sT)^{-1}\int dt\; g_0\; u_{ijkl}(t)$ with 
\be
u_{ijkl}(t)=\int dxdy\;w_i^*w_j^*w_k w_l.
\label{suijkl}
\ee
In the laboratory frame, the Wannier states $w_i(x,y,t)$ of the flat band are superposition of localized wave packets evolving periodically with the period $sT$.	
Indices $i,j, \ldots $ label sites of the effective square lattice which correspond to a unit cell index of the Lieb lattice, cf. Fig.~\ref{regale}. In the course of time evolution different localized wavepackets can overlap in the laboratory frame at different moments of time. The strength $g_0$ of the atom-atom interactions depends on the s-wave scattering length and can be controlled by means of the Feshbach resonance \cite{Chin2010}. Suppose that $g_0$ is periodically modulated in time, i.e.  $g_0(t)=g_0(t+sT)$. The interaction strength $g_0(t)$ can be turned on only for a moment of time when specific Wannier states overlap in the laboratory frame. 
Thus, we can engineer the interaction coefficients $U_{ijkl}$ in the flat band system, Eq.~(\ref{H_fband}), almost at will
which allows one to explore different exotic flat band models.  
Let us analyze what kinds of the models are attainable in the flat band of the Lieb lattice potential presented in Fig.~\ref{eff_pot}(b).

Even if localized wavepackets belonging to Wannier states $w_i$, $w_j$, $w_k$ and $w_l$ overlap in the laboratory frame at a certain moment of time, it does not necessarily mean that the corresponding $u_{ijkl}(t)$ in (\ref{suijkl}) is not zero. An atom which occupies a localized wavepacket is characterized by a quite well defined momentum and if the sum of the momenta of two atoms before and after a collision at $t$ is not conserved, the corresponding $u_{ijkl}(t)$ vanishes. 
If, however, $u_{ijkl}(t)$ does not vanish at a certain time moment $t$, then, we can get the interaction coefficient $U_{ijkl}$ as we wish by choosing an appropriate $g_0(t)$. In the case of the flat band of the Lieb lattice presented in Fig.~\ref{eff_pot}(b), effective selection rules for non-vanishing $u_{ijkl}(t)$ are illustrated in Fig.~\ref{regale}(b). Corners of a symmetrically located rectangle in Fig.~\ref{regale}(b)
correspond to the same position in the Cartesian space $\{x,y\}$ but to four different pairs of the  momenta $\{\pm p_x,\pm p_y\}$ \cite{Sup1}. If at a  certain $t$ four localized wavepackets are at the corners of a certain symmetric rectangle, then we have a guarantee that $u_{ijkl}(t)$ does not vanish, which enables simultaneous hopping of two atoms on the Lieb lattice.  Note that two wavepackets corresponding to the same Wannier state are not necessarily neighbors in the laboratory frame.

To sum up, apart from the simultaneous hopping of pairs of atoms described in Fig.~\ref{regale}, on-site and long-range density-density interactions can be present in the flat band but no density induced tunneling is allowed. Taking into account all possible processes, a general many-body effective Floquet Hamiltonian in the flat band becomes 
\be \label{H_fband_2}
H_F=  \sum_{i} U_i \hat n_i(\hat n_i-1)-\sum_{\{ ijkl \}}J_{ijkl}\hat b^\dagger_i\hat b_j^\dagger\hat b_k\hat b_l,
\ee
where $\hat n_i=\hat b_i^\dagger\hat b_i$. The first sum describes the on-site interactions with the coupling strengths $U_i=U_{iiii}$ while the second sum, with terms proportional to $J_{ijkl}=4U_{ijkl}\small|_{i\ne j}$, is responsible for the long-range density-density interactions and the simultaneous hopping of pairs of atoms. In Fig.~\ref{regale}(c) we illustrate simultaneous hopping of atoms by only two lattice sites and other possible kinds of hopping are shown in \cite{Sup1}. 
Studies of many-body phases of the Lieb model we describe here is beyond the scope of the present letter.


{\bf Conclusions.} 
In this Letter we show that a very simple setting of two oscillating mirrors has a potential for realization of non-equilibrum many-body physics on inseparable lattices with the M\"obius strip geometry. Our system reduces to a time lattice where localized wavepackets are moving along classical resonant orbits. By controlling the periodic motion of the mirrors one is able to design arbitrary lattice geometries. We argue that the effective interactions of the model can be exotic, long-ranged and experimentally tunable. In order to emphasize these peculiar features we focus on a flat band of the Lieb lattice with interaction induced long-distance simultanious hoppings of atomic pairs. Another unique property of our construction is that the 2D time crystalline structures have the geometry of the M\"obius strip.  It is known that the lack of translational symmetry of the M\"obius strip can change the ground state and low energy physics properties of many-body models \cite{Boada2012}. Therefore, our results not only opens up new perspectives for the exploration of interaction induced phenomena, such as exotic superfluids and supersolids on a flat band or the strongly correlated constrained dynamics in the strongly interacting models, but also enable the study of topological effects due to the non-trivial lattice geometry.



{\bf Acknowlegement.}
KG and A. Kuro\'s contributed equally to the present work. This work was supported by the National Science Centre, Poland via Projects No. 2016/20/W/ST4/00314 and No. 2019/32/T/ST2/00413 (KG), QuantERA Programme No. 2017/25/Z/ST2/03027 (A. Kuro\'s),  No.  2018/31/B/ST2/00349 (A. Kosior and KS). KG acknowledges the support of the Foundation for Polish Science (FNP).

\section{Supplemental Material}
In the Supplemental Material, we present details of the classical and quantum analysis of an atom bouncing resonantly between two oscillating mirrors which form a $45^\circ$ wedge. The analysis is particularly convenient in the moving frame of reference, where the effective description of the problem can be reduced to a single particle in a periodic potential. 

In the following sections we first explain how to obtain the tight-binding model that describes interacting particles on the flat band of the Lieb lattice and present all possible effective long-distance pair hopping processes that can be induced by contact interactions between ultra-cold atoms. In the later parts we finally discuss a validity of the effective Hamiltonian and consider higher order corrections to the secular Hamitlonian. 

\section{Unperturbed problem}
Let us first consider the unperturbed problem of static mirrors forming a $90^\circ$ wedge, which is integrable and separable in Cartesian coordinates \cite{Richter1990,wojtkowski1990system}. The corresponding energies $E_x$ and $E_y$ are integrals of motion and it is therefore quite easy to obtain the action-angle variables. The exactly same variables turn out to be very convenient also in the description of the  $45^\circ$ wedge problem, which is not separable but also integrable.
\subsection{Perpendicular mirrors}
The unperturbed Hamiltonian, 
\be
H_0(x,y,p_x,p_y)=\frac{p_x^2}{2}+x+\frac{p_y^2}{2}+y, 
\label{SH0} 	
\ee 
{where $x,y \ge 0$, in the action-angle variables ($I_x,I_y,\theta_x,\theta_y$) depends on the actions only \cite{landau1982mechanics,Lichtenberg1992}
\be
H_0(I_{x},I_{y})=\frac{(3\pi)^{2/3}}{2}\left(I_{x}^{2/3}+I_{y}^{2/3}\right), 
\label{haa}
\ee
where 
\be
  I_{\alpha}=\frac{\left(2 E_{\alpha} \right)^{3/2}}{3\pi}, \quad \theta_{\alpha}=\pi \left( 1-\frac{p_{\alpha}}{\sqrt{2E_{\alpha}}} \right),
\ee
with $\alpha=x,y$. The actions $I_{\alpha}$ are constants of motion and the angles $\theta_{\alpha}$ evolve linearly in time $\theta_{\alpha}=  \Omega_{\alpha}(I_\alpha) t + \theta_{\alpha}(0) \, (\mbox{mod}\, 2\pi)$, where $\Omega_{\alpha}(I_\alpha)=\partial H_0/\partial I_{\alpha}=\pi^{2/3}/(3I_{\alpha})^{1/3}$ are the frequencies of motion of a particle along the $x$ and $y$ directions. The canonical transformation from the action-angle variables to the Cartesian coordinates is given by
\be
\alpha=\frac{(3I_{\alpha})^{2/3}}{2\pi^{4/3}}(2\pi -\theta_{\alpha})\theta_{\alpha},
\ee
and
\be
p_{\alpha}= \left(\frac{3 I_{\alpha}}{\pi^2}\right)^{1/3}(\pi-\theta_{\alpha}).
\ee 
In the Letter we consider the symmetric case where the unperturbed energies $E_x$ and $E_y$ corresponding to the $x$ and $y$ degrees of freedom are equal $E_x=E_y$ and consequently $I_{x}=I_{y}$. In this case the system is classically degenerate and both the frequencies are identical $\Omega_x(I_x)=\Omega_y(I_y)$. We can switch from the variables ($I_{\alpha},\theta_{\alpha}$) to a new set of the action-angle variables ($I_{\pm},\theta_{\pm}$) where for $E_x=E_y$ one of the new frequencies is zero \cite{antonowicz1981freedom}
\bea
\theta_{+}&=&\frac{\theta_x+\theta_y}{2}+\pi h(\theta_x-\theta_y)\sign(2\pi -\theta_x-\theta_y), \cr && \\
\theta_{-}&=&\frac{\theta_y-\theta_x}{2}+\pi h(\theta_x-\theta_y),\\
I_{\pm}&=&I_y \pm I_x,
\eea
where $h(x)$ is the Heaviside step function, $\theta_{+} \in [ 0,2\pi)$ and $\theta_{-} \in [0,\pi)$. The Hamiltonian $H_0$ in the new variables has the form
\be
H_{0}(I_+,I_-)=\frac{1}{2}\left(\frac{3\pi}{2}\right)^{2/3} \left[(I_{+}+I_{-})^{2/3}+(I_{+}-I_{-})^{2/3}\right],
\label{Sup_H0}
\ee
where for $I_x=I_y$ we obtain $I_{+}=2I_x$, $I_{-}=0$ and $\theta_{+}=\Omega_{+}(I_+,I_-)t+\theta_{+}(0)$ while $\theta_{-}=\rm constant$. Indeed, one can easily see that 
\be
\Omega_{+}(I_+,I_-)=\left.\frac{\partial H_0 }{\partial I_+}\right|_{I_{-}=0}=\left(\frac{2\pi^2}{3I_+}\right)^{1/3},
\label{SOmega+}
\ee 
while 
\be
\Omega_{-}(I_+,I_-)=\left.\frac{\partial H_0 }{\partial I_-}\right|_{I_{-}=0}=0. 
\ee
For the sake of completeness, the inverse transformation, i.e. from ($I_{\pm},\theta_{\pm}$) to ($I_{\alpha},\theta_{\alpha}$), reads
\bea
\theta_{x}&=&\theta_{+}-\theta_{-}+2\pi h(\theta_{-}-\theta_{+}),\\
\theta_{y}&=&\theta_{+}+\theta_{-}-2\pi h(\theta_{-}+\theta_{+}-2\pi),\\
I_{x}&=&\frac{1}{2}(I_{+} - I_{-}),\\
I_{y}&=&\frac{1}{2}(I_{+}+I_{-}).
\eea

\subsection{Wedge with an angle $45^\circ$}
The action-angle variables introduced in the previous subsection are useful to identify the topology of the phase space in the case of the wedge with the angle $45^\circ$. Due to the presence of the vertical mirror one should impose extra conditions which are not captured by the definition of the ($I_{x,y},\theta_{x,y}$) and ($I_{\pm},\theta_{\pm}$) variables. Such conditions correspond to the constraint $y\ge x$ which for $I_x=I_y$ reduces to $(\theta_x-\theta_y)(\theta_x+\theta_y-2\pi)\ge 0$ and $\pi-\theta_{+} \ge0$. Moreover, at $y=x$ both the momenta are reversed in the opposite directions, $p_{\alpha}\rightarrow -p_{\alpha}$, which implies the M\"obius strip geometry in the $\{\theta_x,\theta_y\}$ (or $\{\theta_+,\theta_-\}$) space (see Fig.~\ref{moebius_xy} and Fig.~1 in the Letter).

\begin{figure}[h!]
\includegraphics[width=.49\columnwidth]{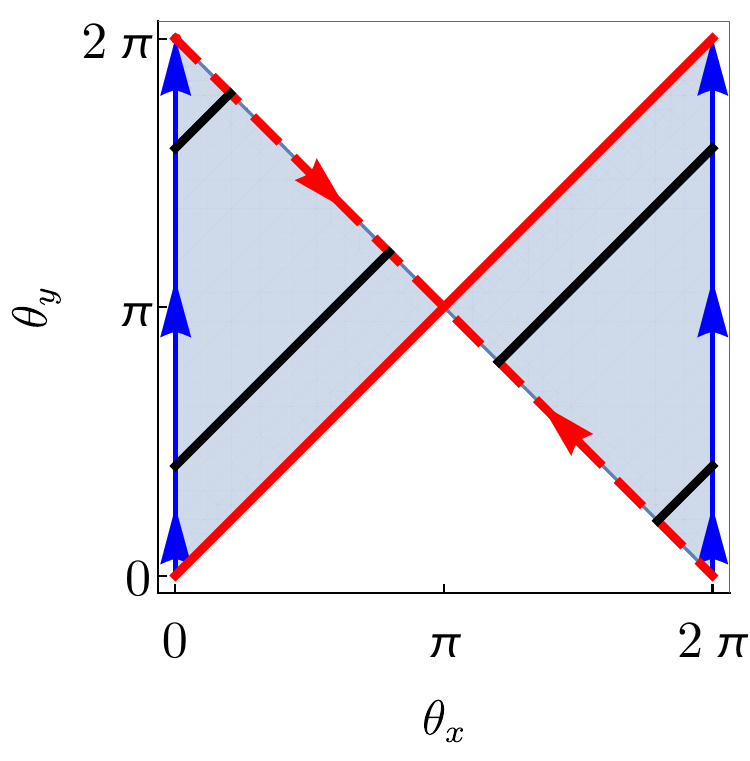}
\includegraphics[width=.49\columnwidth]{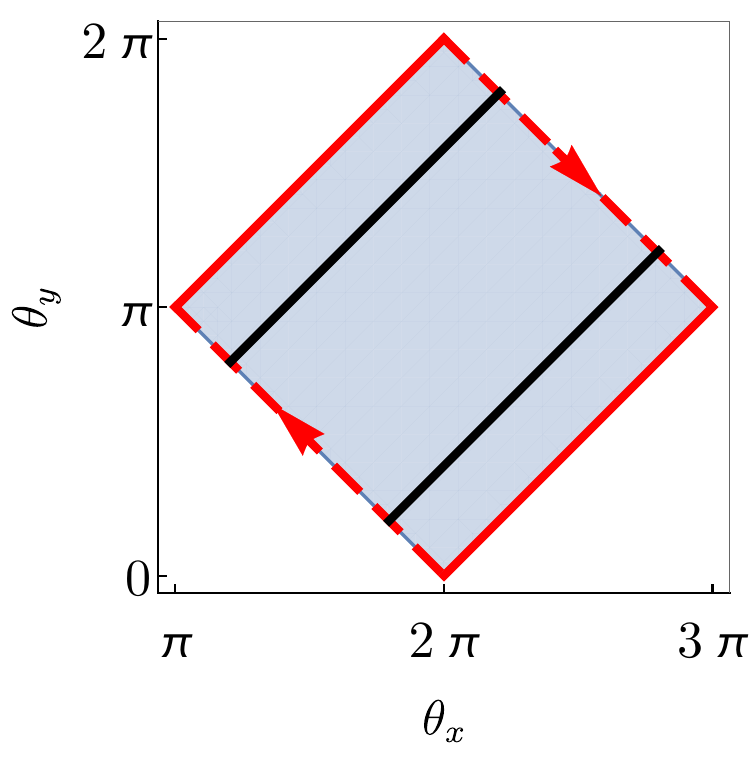}
\caption{M\"obius strip in the $\theta_{x}$ and $\theta_{y}$ variables. Left panel: the condition $y\ge x$ entails the restricted phase space domain $(\theta_x-\theta_y)(\theta_x+\theta_y-2\pi)\ge 0$ introducing new boundaries. The hard boundary (red solid) corresponds to trajectories along with the vertical mirror ($x=y$). The later defines the M\"obius strip. See also Fig.~1 in the Letter for a similar construction in the $\theta_\pm$ variables. Black lines shows an example of a typical trajectory on the M\"obius strip.}
\label{moebius_xy}
\end{figure}

\section{Periodically oscillating mirrors}

Let us turn on oscillations of the mirrors which results in the following Hamiltonian for a particle
\be
H=H_0+F\left[x+f_x(t)\right]+F\left[y-x+f_{y-x}(t)\right], 
\label{hlab1}
\ee
where $H_0$ is given in (\ref{SH0}) and $F$ is a function that models a repulsive potential of the mirrors (in the following we assume the hard wall potential).
The functions $f_{x}(t)$ and $f_{y-x}(t)$ describe oscillations of the mirrors with the period $T=2\pi/\omega$.

In the theoretical description it is convenient to switch from the laboratory frame to the frame oscillating with the mirrors because then the mirrors are fixed and time dependence appears in effective gravitational force. Performing the canonical transformations
$\tilde{x}=x+f_x(t)$, $\tilde{y}=y+f_x(t)+f_{y-x}(t)$, $\tilde{p}_{x}=p_{x}+f_{x}'(t)$,  $\tilde{p}_{y}=p_{y}+f_{x}'(t)+f_{y-x}'(t)$ and dropping the tilde over the variables we obtain
\be
H=H_0+V_{x+y} +V_{y},
\label{Sup_Hfull}
\ee
where $H_0$ is the same as in (\ref{SH0}) and
\bea
V_{x+y}&=&(x + y)\; f_x''(t), \\ 
V_{y}&=&y\; f_{y-x}''(t), 
\eea
with the constraint $y\ge x\ge 0$ coming from the hard wall potential (for a Gaussian shaped  mirror potential see \cite{giergiel2020creating}).

\subsection{Secular Hamiltonian}

When the mirrors oscillate with the frequency $\omega$ we are interested in the motion of a particle in the vicinity of a periodic orbit corresponding to the unperturbed energies $E_x=E_y$. The period of the orbit $2\pi/\Omega_+(I_{+}^{0},I_{-}^{0})$, cf.~(\ref{SOmega+}), is $s$ times longer than the driving period $2\pi/\omega$ where $I_{+}^{0}$ is the resonant value of the action $I_+$ while the resonant value of the other action $I_-=I_{-}^{0}=0$. Let us switch to the frame moving along such an orbit
\bea
\Theta_{+} &=& \theta_{+}-\frac{\omega}{s}t,\\
\Theta_{-} &=& \theta_{-}.
\label{fullaa}
\eea
For actions $I_+$ and $I_-$ close to the resonant values $I_{+}^{0}$ and $I_{-}^{0}=0$, respectively, all variables, i.e. $I_{\pm}$ and $\Theta_\pm$, change slowly. The Cartesian coordinates $x$ and $y$ can be expanded in the Fourier series 
\be
\alpha=\sum_{n=-\infty}^{\infty} c^\alpha_{n}(I_{+},\Theta_{-}) e^{i n (\omega t/s+\Theta_{+}) },
\ee
where 
\bea
c_n^{y}=\begin{cases}
	\frac{I_{+}^{2/3}(2\pi^2+3\pi \Theta_{-}-3\Theta_{-}^2)}{(2^5 3\pi^4)^{1/3}} \  \  \   \  \  \ \  \ \ \  \text{ if } n=0, \\
	-\frac{(3I_{+})^{2/3}(\pi+(-1+(-1)^n)\Theta_{-})}{2^{2/3} n^2 \pi^{7/3}}  \  \ \text{ if } n\neq 0, \\
\end{cases}\,
\eea
and
\bea
c_n^{x}=\begin{cases}
	\frac{I_{+}^{2/3}(2\pi^2-3\pi \Theta_{-}+3\Theta_{-}^2)}{(2^5 3\pi^4)^{1/3}} \ \  \   \ \ \ \ \  \ \ \  \  \ \ \ \  \  \ \ \  \ \ \  \text{ if } n=0, \\
	-\frac{\pi(1+e^{2i n \Theta_{-}})-e^{i n \Theta_{-}}(\pi+(-1+(-1)^n)\Theta_{-})}{2^{2/3} n^2 \pi^{7/3}(3I_{+})^{-2/3}e^{i n \Theta_{-}}}  \   \text{ if } n\neq 0. \\
\end{cases}\,
\label{classElems}
\eea
In the action-angle variables, the unperturbed part of the Hamiltonian is given by $H_0-\omega I_+/s$ with $H_0$ like in (\ref{Sup_H0}) and the perturbations
\bea
V_{x+y}&=&f_x''(t)\sum_{n=-\infty}^{\infty} (c^{x}_{n}+c^{y}_{n}) e^{i n (\omega t/s+\Theta_{+}) },
\\
V_{y}&=&
f_{y-x}''(t)\sum_{n=-\infty}^{\infty} c^{y}_{n} e^{i n (\omega t/s+\Theta_{+}) }.
\eea
As an example let us consider the following driving $f_{y-x}(t)=f_{x}(t)=\lambda/(k^2\omega^2)\cos(k \omega t+\phi)$, where $k$ is an integer number and $\phi$ an arbitrary phase. Assuming the resonance condition, i.e. $\omega=s\Omega_{+}(I_{+}^{0},I_{-}^{0})$ where $s$ is an even integer number, we can carry out averaging of the Hamiltonian (\ref{Sup_Hfull}) over time keeping all dynamical variables fixed. However, we should remember that when for fixed $\Theta_\pm$, the position variable in the lab frame $\theta_+=\Theta_++\omega t/s$ reaches $\pi$ we have to switch $\Theta_\pm\rightarrow \pi-\Theta_\pm$. The resulting effective potential reads
\be
\langle V_{x+y} \rangle_t= \frac{2\lambda}{k^2 \omega^2} \cos(ks \Theta_{+}+\phi)\cos(ks \Theta_{-} ) ,
\ee
and
\be
\langle V_{y} \rangle_t= \frac{\lambda}{k^2 \omega^2} \cos(ks \Theta_{+}+\phi) .
\ee
Performing the Taylor expansion of $H_0(I_+,I_-)$ around the resonant values $I_{\pm}^{0}$ of the actions, we can express the entire effective Hamiltonian as follows (with a constant term omitted)
\bea
H_{\rm eff} &\approx&  \frac{P_{+}^2+P_{-}^2}{2 m_{\rm eff}}+\frac{2\lambda}{k^2 \omega^2} \cos(ks \Theta_{+}+\phi)\cos(ks \Theta_{-}) +\cr &&
\frac{\lambda}{k^2 \omega^2} \cos(ks \Theta_{+}+\phi),
\label{heff_sup}
\eea
with the identification of the points $\{\Theta_{+}=\pi,\Theta_{-}\}=\{\Theta_{+}=0,\pi-\Theta_{-}\}$, where 
\be
m_{\rm eff}^{-1}=\left.\frac{\partial^2 H_0(I_{+},I_{-})}{\partial I_{\pm}^2}\right|_{I_{\pm}=I_{\pm}^{0}},
\ee 
and $P_{\pm}=I_{\pm}-I_{\pm}^{0}$. The same Hamiltonian (\ref{heff_sup}), but in the ($\Theta_x, \Theta_y, I_x, I_y$) variables has the form 
\bea
H_{\rm eff}&\approx&\frac{P_x^2+P_y^2}{2m_{0}}+\frac{\lambda}{k^2\omega^2}\cos \left(\frac{ks}{2}(\Theta_x+\Theta_y)+\phi\right)
\cr && + \frac{\lambda}{k^2\omega^2}\left[\cos(ks\Theta_x+\phi)\right. \left.+ \cos(ks\Theta_y+\phi)\right] , 
\cr &&
\label{heff_sup2}
\eea
with the constraint $\Theta_y(2\pi-\Theta_y) \geq \Theta_x (2\pi -\Theta_x)$, where $P_{x,y}=I_{x,y}-I_{x,y}^{0}$ and $m_{0}=(\partial^2 H_0(I_x,I_y)/\partial I_{x,y}^2)|_{I_{x,y}=I_{x,y}^{0}}$.

\section{Lieb lattice}

\subsection{Tight-binding approximation}
\label{sectight}
If the mirrors, that form the wedge with the angle $45^\circ$, oscillate according to (cf. Eq.~(1) in the Letter) 
\bea
f_x(t)&=&-\frac{\lambda_1}{\omega^2}\cos(\omega t)-\frac{\lambda_2}{4\omega^2}\cos(2\omega t),
\\ 
f_{y-x}(t)&=&\frac{\lambda_3}{4\omega^2}\cos(2\omega t+\phi),
\eea 
then, for $\lambda_2/\lambda_1=4$, $\lambda_3/\lambda_2=1.62$ and $\phi=\pi/4$, the classical effective Hamiltonian,
\bea
H_{\rm eff}&=&-\frac{P_{-}^2+P_{+}^2}{2|m_{\rm eff}|}- \frac{\lambda_2}{2\omega^2} \cos \left(2s\Theta_{+} \right) \cos \left(2s\Theta_{-}\right) 
\cr &&
- \frac{2\lambda_1}{\omega^2}\cos(s\Theta_{+})\cos(s\Theta_{-}) +\frac{\lambda_3}{4 \omega^2}\cos \left(2s\Theta_{+}+\phi\right), 
\cr &&
\label{Sh_effLieb}
\eea 
describes a particle in the Lieb lattice potential on a M\"obius strip which is presented in Fig.~2(b) in the Letter. 

In order to reduce the quantum description of the system to the tight-binding model, Eq.~(4) in the Letter, we perform the quantum secular approximation. First we define the basis of antisymmetric states 
\be
\psi_{nm}(x,y)\propto\phi_n(x)\phi_m(y)-\phi_m(x)\phi_n(y),
\label{antisym_basis}
\ee
with $n>m$, where $\phi_n$ are eigenstates of the 1D problem of a particle bouncing on a static mirror. The basis states $\psi_{nm}(x,y)$ fulfill the proper boundary conditions on the mirrors. Next we switch to the rotating frame by means of the unitary transformation $e^{i(\hat m+\hat n)\omega t/s}$ and neglect time-oscillating terms which leads to the effective quantum Hamiltonian. Eigenenergies of the effective Hamiltonian form energy bands and we restrict to the Hilbert subspace of the first three bands. In order to define the Wannier states basis in such a subspace we define the plane wave representation of the basis states 
\be
\psi_{nm}(\Theta_x,\Theta_y)\propto \phi_n(\Theta_x)\phi_m(\Theta_y)-\phi_m(\Theta_x)\phi_m(\Theta_y),
\ee
where $\phi_n(\Theta_x)=\la\Theta_x|\phi_n\ra\propto \sin(n\Theta_x)$ and $\phi_m(\Theta_y)=\la\Theta_y|\phi_m\ra\propto \sin(m\Theta_y)$, and diagonalize the operators $e^{i\Theta_x}$ and $e^{i\Theta_y}$ within the subspace. The eigenstates of these operators are the Wannier states $W_{i,\beta}$, where $i$ is a unit cell index, and $\beta=0, \pm$ is a intra cell index, cf. Fig.~2(b) of the Letter. The Wannier states are localized wavepackets $W_{i,\beta}(x,y,t)$ which are moving along resonant orbits in the laboratory  frame with the period $sT$. When we expand the bosonic field operator in the series of annihilation operators $\hat a_{i,\beta}$ which annihilate a boson in the Wannier states, 
\be
\hat \psi(x,y,t)\approx\sum_{i,\beta}W_{i,\beta}(x,y,t)\;\hat a_{i,\beta}, 
\label{Shatpsi}
\ee
we obtain the effective Hamiltonian (which is actually the Floquet Hamiltonian for non-interacting bosons) in the tight-binding form, Eq.~(4) in the Letter, i.e.
\bea
 H_F&=&\frac{1}{sT}\int_0^{sT}dt\int dxdy\;\hat\psi^\dagger\left(H-i\partial_t\right)\hat\psi
\cr &\approx& -J_1 \sum_{i,\beta=\pm}\hat a_{i,0}^\dagger \hat a_{i,\beta}-J_2 \sum_{\la ij\ra,\beta=\pm}\hat a_{i,0}^\dagger \hat a_{j,\beta} +\mbox{H.c.},
\cr &&
\label{Stb}
\eea
where we omitted constant terms. 
Single-particle spectrum of the tight-binding Hamiltonian (\ref{Stb}) is shown in Fig.~\ref{Sup_spectrum} and indicates the presence of three energy bands where the middle one is flat. 

We are interested in the flat band physics and in order to derive the tight-binding model restricted to the flat band we again perform diagonalization of the operators $e^{i\Theta_x}$ and $e^{i\Theta_y}$ but this time in the Hilbert subspace restricted to the eigenstates that belong to the flat band. The diagonalization results in a new set of Wannier states $w_i$ which, for $J_1/J_2\gg 1$, are either nearly identical with the former Wannier states $W_{i,\beta}$ or are  superposition of two states $W_{i,+}$ and $W_{i,+}$, cf. Fig.~3(a) in the Letter.

\begin{figure}[t]
\includegraphics[width=.85\columnwidth]{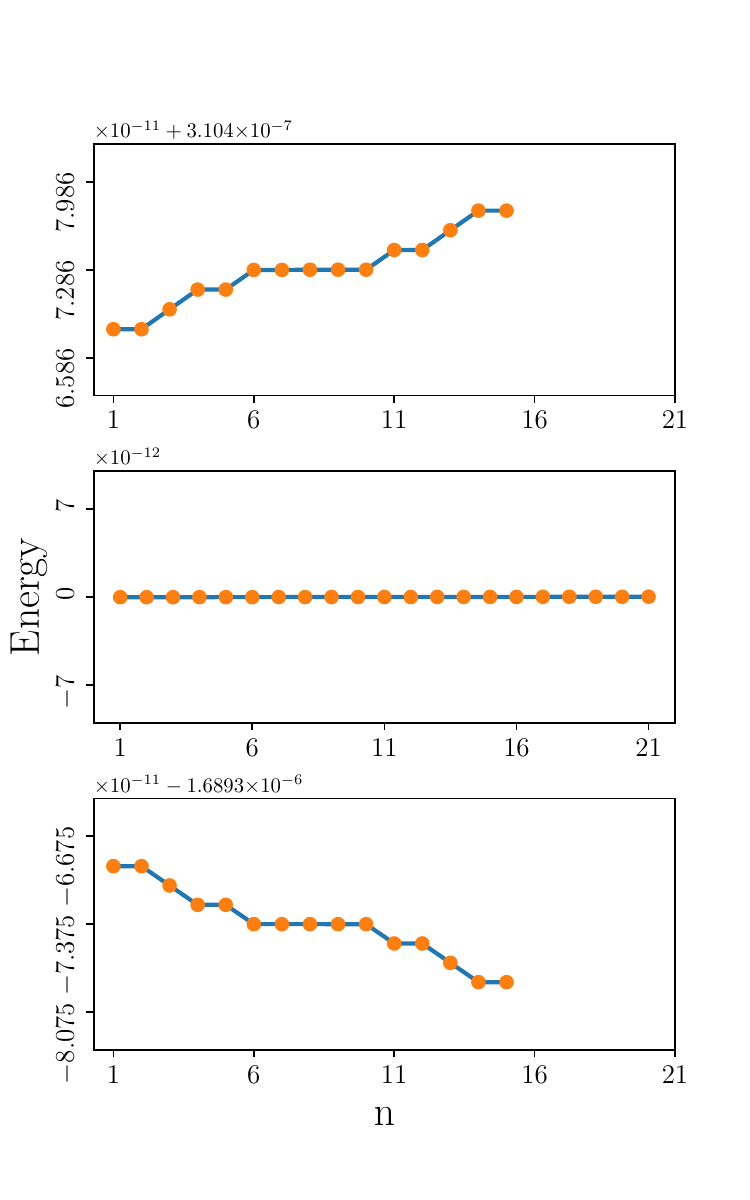}
\caption{
Eigenenergies of the tight-binding Hamiltonian (\ref{Stb}) for a single particle in the Lieb lattice in the case of  $s=6$, cf. (\ref{Sh_effLieb}). Three bands are formed where the middle one is flat. Top panel shows the upper band, middle panel the flat middle band and bottom panel the lower band. Ranges in the vertical axes are the same in all panels which allows us to demonstrate how flat the middle band is as compared to the band widths of the upper and lower bands. Note that the number of energy levels of the flat band is greater by $s=6$ because lattice sites close to the edge of the M\"obius strip belong to the Hilbert space of the flat band only.
}
\label{Sup_spectrum}
\end{figure}

\begin{figure}[t]
\includegraphics[width=.99\columnwidth]{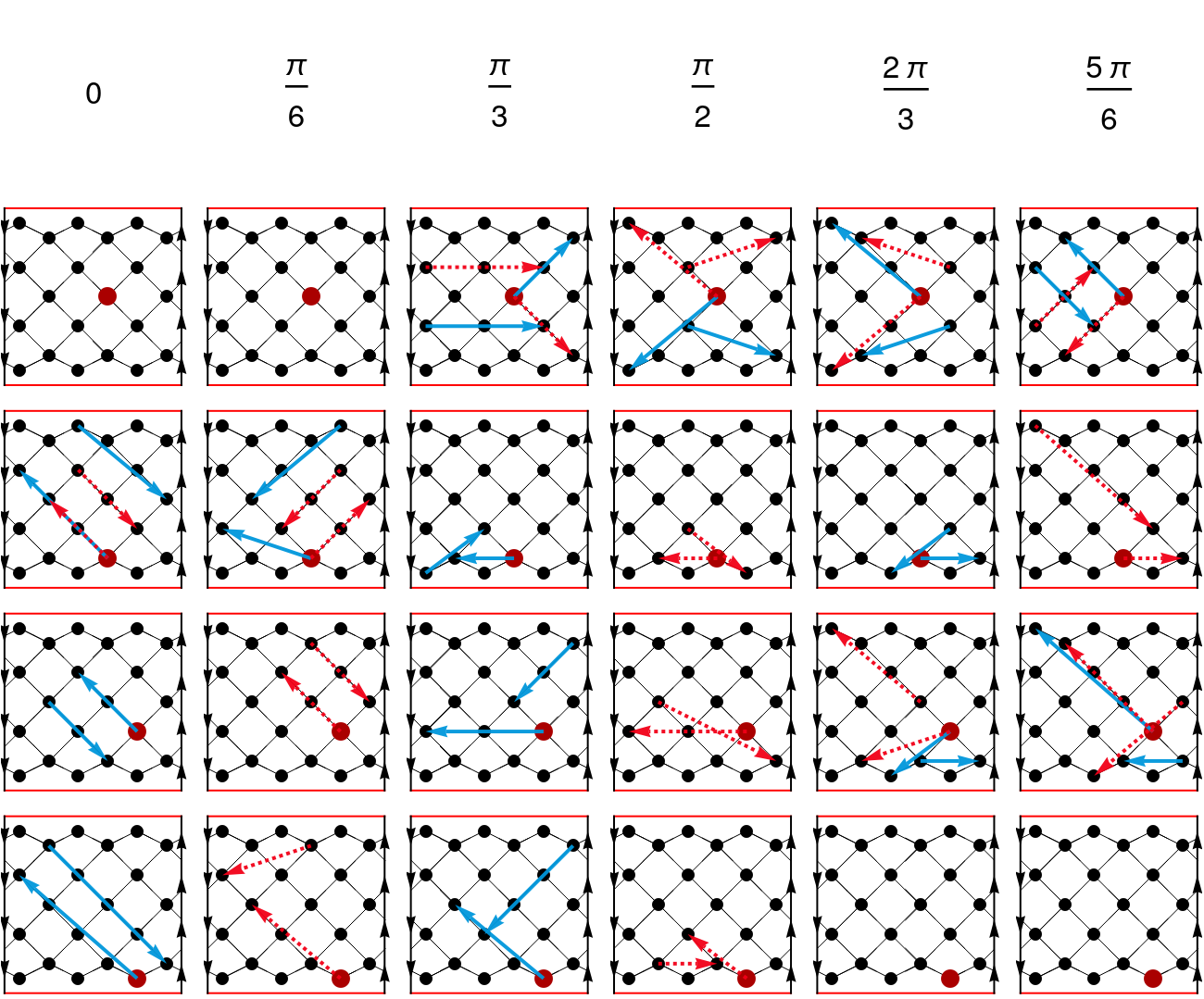}
\caption{Panels present all possible pair hopping between sites of the flat band of the Lieb lattice in the case of $s=6$, cf. (\ref{Sh_effLieb}). Dots denote the Wannier states $w_i$ of the flat band. Each column corresponds to a different moment of time indicated by the value of $\omega t$ on the top of the figure. At different $\omega t$, different pair hopping are possible, i.e. different $u_{ijlk}(t)$ in (\ref{Ssuijkl}) do not vanish. In each row different representative initial sites (indicated by brown dots) where one atom of a hopping pair is located are considered. Note that the Lieb lattice has the M\"obius strip geometry and in order to glue together the left and right sides of each square, one has to first twist it so that the arrows of the both sides of a square point in the same direction. The interaction structure is repeated in the second part of the period $(\pi,2\pi)$. Empty panels correspond to a situation when only two wavepackets meet at some moment of time. In this case, there is no tunneling in the flat band.  
}
\label{Sselectionrules}
\end{figure} 

If the contact interaction between bosons are present and the interaction energy per particle is much smaller than the energy gaps between the flat band and the adjacent bands, to describe the flat band physics we may truncate the bosonic field operator to the sum of the annihilation operators $\hat b_i$ that 
annihilate a boson in the new Wannier states $w_i$, i.e. $\hat \psi(x,y,t)\approx\sum_{i=1}^{s(s+1)/2}w_i(x,y,t)\hat b_i$. It allows us to obtain the desired tight-binding model (Eq.~(5) in the Letter) which describes dynamics of interacting bosons in the flat band, i.e.
\bea
 H_F&=&\frac{1}{sT}\int_0^{sT}dt\int dxdy\;\hat\psi^\dagger\left(H-i\partial_t+\frac{g_0}{2}\hat\psi^\dagger\hat\psi\right)\hat\psi
\cr &\approx&\sum_{ijkl}U_{ijkl}\hat b^\dagger_i\hat b_j^\dagger\hat b_k\hat b_l+{\rm const},
\label{Stbw}
\eea
where 
\be
U_{ijkl}=\int _0^{sT}\frac{dt}{sT}\; g_0\; u_{ijkl}(t),
\ee
with 
\bea
u_{ijkl}(t)&=&\int dxdy\;w_i^*(x,y,t)\;w_j^*(x,y,t)
\cr && \times  w_k(x,y,t)\;w_l(x,y,t).
\label{Ssuijkl}
\eea

The interaction coefficients $U_{ijkl}$ in (\ref{Stbw}), which actually determine hopping of pairs of bosons in the Lieb lattice, depend on the interaction strength $g_0$ which is proportional to the s-wave scattering length of ultra-cold atoms bouncing between the mirrors. Feshbach resonances allows one to change s-wave scattering by means of an external magnetic field. If $g_0$ is changing periodically in time, $g_0(t+sT)=g_0(t)$, then one can control which coefficients $U_{ijlk}$ are significant and which negligible because different Wannier states overlap in the laboratory frame in different moments of time. However, even if four Wannier states  $w_i$, $w_j$, $w_k$ and $w_l$ overlap at certain moment of time $t$, the coefficient $u_{ijkl}(t)$ in (\ref{Ssuijkl}) can still vanish and consequently  the corresponding $U_{ijkl}$ will be zero. The Wannier states $w_i$ consist of a single or two localized wavepackets $W_{i,\beta}$. An atom in a localized wavepacket $W_{i,\beta}$ is characterized by quite well defined momentum. If two atoms occupying different wavepackets collide at time moment $t$, then the coefficient $u_{ijkl}(t)$ does not vanish if the sum of the momenta of the atoms before and after the collision is conserved. It leads to simple selection rules for hopping of pairs of atoms in the Lieb lattice on a M\"obius strip which are explained in Fig.~4 of the Letter. In Fig.~\ref{Sselectionrules} we illustrate all pair hopping which are possible in the Lieb lattice on the M\"obius strip in the case of $s=6$. At different moments of time wavepackets belonging to different Wannier states $w_i$ overlap and different coefficients $u_{ijkl}(t)$ are non-zero. 

\subsection{Validity of the effective many-body Hamiltonian}

We have reduced description of the periodically driven many-body system to the effective Hamiltonian (\ref{Stbw}). The validity of this Hamiltonian requires the interaction energy per particle to be much smaller than the energy gaps between the flat band and the neighboring bands of the tight-binding Hamiltonian (\ref{Stb}) which can be easily fulfilled. However, the interactions between bosons can also couple the resonant subspace spanned by the Wannier states $W_{i,\beta}$, cf. (\ref{Shatpsi}), to the complementary  Hilbert subspace what is neglected in our description. On a very long time scale it may lead to heating of the system because a generic periodically driven many-body system is expected to eventually heat up to the infinite temperature state unless it is integrable. While the analysis of this problem is beyond the scope of the present Letter, we can refer to the results obtained for a similar problem of bosons bouncing resonantly on an oscillating mirror in the 1D case. The Bogoliubov approach \cite{Kuros2020} and the truncated Wigner approximation \cite{Wang2020} do not reveal any signature of heating of the system for thousands of the periods of the mirror oscillation which is by far longer than it is required to perform the experiment.

\subsection{Analysis of corrections to the secular Hamiltonian}

Here, we analyze corrections to the quantum secular Hamiltonian. As an example, let us consider time periodic driving where in (\ref{Sup_Hfull}), $V_{x+y}=0$ and $V_{y}=\lambda y\cos(s\omega t)$ (the presence of $V_y$ is crucial in our analysis because this term couples the spatial degrees of freedom of the particle). Starting with the antisymmetric basis (\ref{antisym_basis}) and switching to the moving frame with the help of the unitary transformation $e^{i(\hat n_x+\hat n_y)\omega t}$ we obtain the Hamiltonian of the particle bouncing between the oscillating mirrors in the form
\bea
\la m_x,m_y|H(t)|n_x,n_y\ra&=&E_{n_x,n_y}\delta_{m_x,n_x}\delta_{m_y,n_y} \cr
&& +\la m_x,m_y|y(t)|n_x,n_y\ra
\cr && \times \lambda \cos(s\omega t),
\label{matrixelementH(t)}
\eea
where $E_{n_x,n_y}$ are eigenvalues of the unperturbed Hamiltonian in the moving frame, $H_0-\omega (\hat n_x+\hat n_y)$, and 
\bea
\la m_x,m_y|y(t)|n_x,n_y\ra&=&\la m_x,m_y|y|n_x,n_y\ra 
\times e^{-i(n-m)\omega t},
\cr &&
\label{matrixalpha}
\eea
with $n=n_x+n_y$ and $m=m_x+m_y$. 

In order to calculate the quantum secular Hamiltonian and analyze corrections to it, let us apply the Magnus expansion (see e.g. \cite{Blanes2010}),
\bea
H^{(0)} &=& \frac{1}{T}\int\limits_{0}^{T} d t_1 H(t_1),
\label{appen_fristMagnus}
\\
H^{(1)} &=& \frac{1}{2Ti}\int\limits_{0}^{T} d t_1 \int\limits_{0}^{t_1} d t_2 \left[H(t_1), H(t_2) \right],
\label{appen_secondMagnus}
\eea
where $H(t)$ is given in (\ref{matrixelementH(t)}).
The first term of the Magnus series, Eq.~(\ref{appen_fristMagnus}), corresponds to the quantum secular Hamiltonian used in the Letter,
\bea
\la m_x,m_y|H^{(0)}|n_x,n_y\ra&=&E_{n_x,n_y}\delta_{m_x,n_x}\delta_{m_y,n_y} \cr
&& +\lambda \la m_x,m_y|y|n_x,n_y\ra
\cr && \times \frac{1}{2}(\delta_{m,n-s}+\delta_{m,n+s}).
\cr &&
\label{zerH_F}
\eea
We restrict to the resonant Hilbert subspace where $|n_{x,y}-n_0|\ll n_0$ and $|m_{x,y}-n_0|\ll n_0$ with $n_0$ being the resonant quantum number (i.e. quantum analogue of the classical resonant action $I_s$).
The second term in the Magnus series, $H^{(1)}$, has been omitted in the description of the system and we are going to show that it is negligible if we choose properly the parameters of the system. 

Analyzing the classical secular Hamiltonian (\ref{heff_sup}) [or (\ref{heff_sup2})] it becomes evident that when we switch from $n_0$ to $n_0'$ but at the same time multiply $\lambda$ by $n_0^2/{n_0'}^2$, we obtain exactly the same dynamics because the new and old secular Hamiltonians differ by a multiplicative constant only. Indeed, $m_{\rm eff}\propto n_0^{4/3}$, $\omega\propto n_0^{-1/3}$, $\la m_x,m_y|y|n_x,n_y\ra\propto n_0^{2/3}$ and if we assume $\lambda \propto n_0^{-2}$, then for any $n_0$ we get the same dynamics. 

The commutator in (\ref{appen_secondMagnus}) consists of the first and second order contributions in $\lambda$. The first order one results in
\begin{widetext}
\bea
C_1&=&\lambda\langle m_x,m_y|y| n_x,n_y\rangle\sum_{j=\pm1} \frac{E_{n_x,n_y}-E_{m_x,m_y}}{2\omega(n-m+j s)}
\cr 
&\approx&\lambda\langle m_x,m_y|y| n_x,n_y\rangle\sum_{j=\pm1} \frac{(n_x-n_0)^2+(n_y-n_0)^2-(m_x-n_0)^2-(m_y-n_0)^2}{4m_{\rm eff}\omega(n-m+j s)}.
\label{magnuslambda}
\eea
\end{widetext}
There is no small denominator problem in Eq.~(\ref{magnuslambda}) because these Magnus terms have been obtained with the assumption $n-m\pm s\ne 0$ otherwise they are zero. The term $C_1$ is a negligible correction to the secular Hamiltonian (\ref{zerH_F}) if we choose sufficiently large $n_0$ and assume that $\lambda\propto n_0^{-2}$. Indeed, the matrix elements of the secular Hamiltonian scale with $n_0$ like $n_0^{-4/3}$, while $C_1\propto n_0^{-7/3}$ and can be omitted. To estimate $C_1$ we have assumed that $(n_{x,y}-n_0)^2\le\rm constant$, i.e. the matrix of the secular Hamiltonian is truncated in the same way independently of $n_0$ because for $n_0\gg 1$ the dynamics is the same if we use the scaling $\lambda\propto n_0^{-2}$.

The second order term in $H^{(1)}$ reads 
\begin{widetext}
\bea
C_2&=&\left|\frac{\lambda^2}{8\omega}\sum\limits_{k_x,k_y}\sum_{j=\pm 1} \langle m_x,m_y|y| k_x,k_y\rangle\langle k_x,k_y|y| n_x,n_y\rangle \right.
\cr && 
\times \left\{\left(\frac{1}{n-k-j s}+\frac{1}{m-k+j s}\right)\delta_{n,m+j 2s} \right. 
+\left(\frac{1}{n-k+j s}+\frac{1}{m-k+js}\right)\delta_{n,m}
\label{Magnuslambda2a}
\\ && \left. + \left(\frac{2}{k-n+ s} +\frac{2}{k-n-s}\right)\delta_{m,k+j s} 
 \left. + \left(\frac{2}{k-m+ s}+\frac{2}{k-m- s}\right)\delta_{n,k+j s}\right\}\right|
 \label{Magnuslambda2b}
 \\ &<&
 \frac{3\lambda^2}{\omega}\langle m_x,m_y|y^2| n_x,n_y\rangle,
\label{Magnuslambda2}
\eea
\end{widetext}
where, similarly like in (\ref{magnuslambda}), only terms with non-vanishing denominators in (\ref{Magnuslambda2a})-(\ref{Magnuslambda2b}) contribute to the sum. To obtain the estimate (\ref{Magnuslambda2}) we have neglected the dependence on $k$ of the terms in (\ref{Magnuslambda2a})-(\ref{Magnuslambda2b}) which constitutes a very rough upper bound of $C_2$. When $n_0$ increases, we get the following bound: $C_2<Cn_0^{-7/3}$, where $C$ is a constant. Thus, both $C_1$ and $C_2$ can be neglected in the large $n_0$ limit and there is no correction to the secular Hamiltonian from the leading Magnus terms.

\section*{References}
\bibliography{ref_06_2021.bib}

\begin{thebibliography}{74}
\bibitem[{foo()}]{footnote1}
\bibinfo{note}{We use the gravitational units, but assume that the
  gravitational acceleration is given by $g/\sqrt{2}$.}
\end{thebibliography}
\end{document}